\RequirePackage[mathlines]{lineno} 
\documentclass[a4paper,11pt]{article}

\usepackage{jheppub} 
\usepackage{overpic}
\usepackage{amssymb}
\usepackage{amsmath}
\usepackage[T1]{fontenc} 
\usepackage{amsmath}
\usepackage{graphicx}
\usepackage{subfigure}
\usepackage{epstopdf}
\usepackage{rotating}
\usepackage{adjustbox} 

\usepackage{threeparttable}
\usepackage{multirow}
\usepackage{subfigure}
\usepackage{float}
\usepackage{verbatim}

\let\oldequation\equation
\let\oldendequation\endequation
\renewenvironment{equation}
  {\linenomathNonumbers\oldequation}
  {\oldendequation\endlinenomath}

\begin{document}

\title{\bf \boldmath
Search for the radiative decays $D^+\to\gamma\rho^+$ and $D^+\to\gamma K^{*+}$
}

\collaborationImg{\includegraphics[height=4cm,angle=0 ]{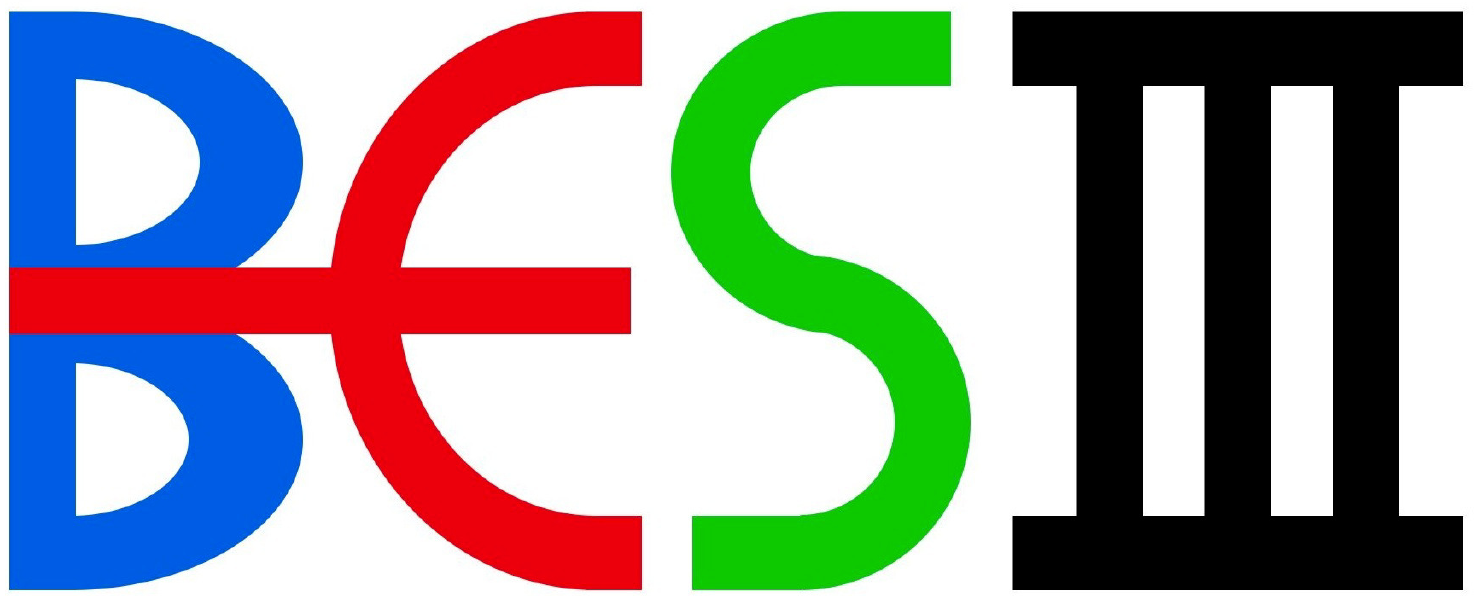}}
\collaboration{The BESIII Collaboration}
\emailAdd{besiii-publications@ihep.ac.cn}

\abstract{
   We search for the radiative decays $D^{+} \to \gamma \rho^+$ and $D^{+} \to \gamma K^{*+}$
  using 20.3~fb$^{-1}$ of $e^+e^-$ annihilation data collected at the center-of-mass energy $\sqrt{s}=3.773$ GeV by the BESIII detector operating at the BEPCII collider.
  No significant signals are observed, and the upper limits on the branching fractions of
  $D^{+} \to \gamma \rho^+$ and $D^{+} \to \gamma K^{*+}$  at 90\% confidence level are set to be $1.3\times10^{-5}$ and $1.8\times10^{-5}$, respectively.
}

\maketitle
\flushbottom

\section{Introduction}

\begin{figure}[htpb]
  \centering
  \includegraphics[width=0.6\textwidth]{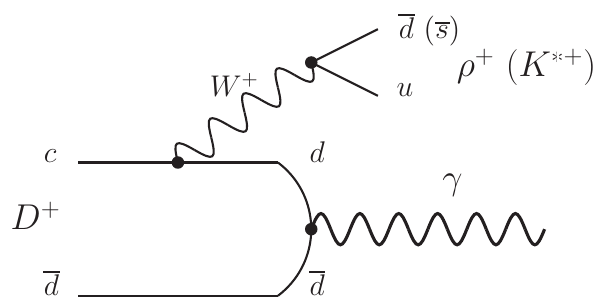}
  \caption{Feynman diagrams for $D^+ \to \gamma \rho^+$ and $D^+ \to \gamma K^{*+}$.}
  \label{fig:feynman}
\end{figure}

The radiative decays $D^+\to\gamma \rho(770)^+$ and $D^+\to\gamma K^*(892)^+$ can be mediated via Feynman diagrams as shown in Fig~\ref{fig:feynman}.
The study of weak radiative decays of charmed hadrons allows for direct observation of the dominant contributions from long-range effects that are independent of new physics, thereby testing theoretical calculations of non-perturbative QCD~\cite{Burdman:1995te,Fajfer:2015zea}.
Experimental measurements of the branching fractions of charmed meson decays are important to test QCD-based calculations of long-distance dynamics.
Previously, the radiative decays
$D^0\to\gamma \bar K^{*0}$,
$D^0\to\gamma\rho^0$,
$D^0\to\gamma\omega$,
and $D^0\to\gamma\phi$
have been measured by Belle~\cite{Belle:2003vsx,Belle:2016mtj}, BaBar~\cite{BaBar:2008kjd}, and CLEO~II~\cite{CLEO:1998mtp}.
However, the radiative decays of $D^+\to\gamma \rho^+$ and $D^+\to\gamma K^{*+}$ have not been studied yet.
Throughout this paper, charge conjugations are implied.

We report the first search for the radiative decays $D^{+} \to \gamma \rho^+$ and $D^{+} \to \gamma K^{*+}$
by analyzing a data sample collected in $e^+e^-$ annihilations at the center-of-mass energy of 3.773 GeV with the BESIII detector, corresponding to an integrated luminosity of 20.3~fb$^{-1}$~\cite{BESIII:2024lbn}.
Various theoretical approaches predict the branching fraction of the Cabibbo-suppressed process $D^+\to \gamma \rho^+$~\cite{Fu:2018yin,Khodjamirian:1995uc,deBoer:2017que,Fajfer:1997bh,Fajfer:1998dv,Burdman:1995te} up to a value of $10^{-5}$, which is accessible with the BESIII data sample.
In addition, it is naively expected that the branching fraction of the doubly Cabibbo-suppressed process $D^+\to \gamma K^{*+}$ is one order of magnitude lower than that of $D^+\to \gamma \rho^+$.

\section{Data and Monte Carlo simulation}

The BESIII detector~\cite{BESIII:2009fln} records symmetric $e^+e^-$ collisions
provided by the BEPCII storage ring~\cite{Yu:2016cof}
in the center-of-mass energy range from 1.85 to 4.95~GeV,
with a peak luminosity of $1.1 \times 10^{33}\;\text{cm}^{-2}\text{s}^{-1}$
achieved at $\sqrt{s} = 3.773\;\text{GeV}$.
BESIII has collected large data samples in this energy region~\cite{BESIII:2020nme,lu2020online,zhang2022suppression}. The cylindrical core of the BESIII detector covers 93\% of the full solid angle and consists of a helium-based
multilayer drift chamber~(MDC), a plastic scintillator time-of-flight
system~(TOF), and a CsI(Tl) electromagnetic calorimeter~(EMC),
which are all enclosed in a superconducting solenoidal magnet
providing a 1.0~T magnetic field.
The solenoid is supported by an
octagonal flux-return yoke with resistive plate counter muon
identification modules interleaved with steel.
The charged-particle momentum resolution at $1~{\rm GeV}/c$ is
$0.5\%$, and the
${\rm d}E/{\rm d}x$
resolution is $6\%$ for electrons
from Bhabha scattering. The EMC measures photon energies with a
resolution of $2.5\%$ ($5\%$) at $1$~GeV in the barrel (end cap)
region.
The time resolution in the TOF barrel region is 68~ps, while
that in the end cap region was 110~ps.
The end cap TOF system was upgraded in 2015 using multigap resistive plate chamber technology, providing a time resolution of 60~ps, which benefits 85\% of the data used in this analysis
~\cite{li2017study,guo2017study,CAO2020163053}.

Simulated samples have been produced with the {\sc geant4}-based~\cite{GEANT4:2002zbu} Monte Carlo (MC) package. It includes the geometric description of the BESIII detector and the
detector response, and is used to determine the detection efficiency and to estimate the backgrounds.
The simulation includes the beam-energy spread and initial-state radiation in the $e^+e^-$
annihilations modeled with the generator {\sc kkmc}~\cite{Jadach:2000ir,Jadach:1999vf}.
The inclusive MC samples consist of the production of $D\bar{D}$ pairs,
the non-$D\bar{D}$ decays of the $\psi(3770)$, the initial-state radiation
production of the $J/\psi$ and $\psi(3686)$ states, and the
continuum processes.
The known decay modes are modelled with {\sc evtgen}~\cite{Ping:2008zz,Lange:2001uf} using the branching fractions taken from the
Particle Data Group~\cite{ParticleDataGroup:2024cfk}, and the remaining unknown decays of the charmonium states are
modeled by {\sc lundcharm}~\cite{Chen:2000tv}. Final-state radiation is incorporated using the {\sc photos} package~\cite{Richter-Was:1992hxq}.

The exclusive MC samples of the radiative decays $D^{+} \to \gamma \rho^+$ and $D^{+} \to \gamma K^{*+}$ are generated by taking into account the helicity amplitudes, the same ones as used for the Belle and Babar measurements~\cite{BaBar:2008kjd,Belle:2016mtj}.

\section{Method}
At $\sqrt s=3.773$~GeV, the $D^+D^-$ pairs are produced without any accompanying hadrons,
thereby offering a clean environment to investigate hadronic $D$ decays with the double-tag~(DT) method~\cite{MARK-III:1985hbd,MARK-III:1987jsm}.
The single-tag~(ST) $D^-$ candidates are selected by reconstructing a $D^-$ in the hadronic decay modes:
$D^-\to K^{+}\pi^{-}\pi^{-}$, $K^0_{S}\pi^{-}$, $K^{+}\pi^{-}\pi^{-}\pi^{0}$, $K^0_{S}\pi^{-}\pi^{0}$,
$K^0_{S}\pi^{+}\pi^{-}\pi^{-}$, and $K^{+}K^{-}\pi^{-}$.
Events in which a signal candidate is reconstructed in the presence of an ST $D^-$ meson
are referred to as DT events.
The product branching fraction of the signal decay is determined by
\begin{equation}
  \label{eq:br}
  {\mathcal B}_{{\rm sig}} = N_{\rm DT}/(N^{\rm tot}_{\rm ST}\cdot\epsilon_{{\rm sig}}\cdot{\mathcal B}_{{\rm sub}}),
\end{equation}
where
$N^{\rm tot}_{\rm ST}=\sum_i N_{{\rm ST}}^i$ and $N_{\rm DT}$
are the total yields of the ST and DT candidates in data, respectively.
The ${\mathcal B}_{{\rm sub}}$ are the product of the subdecay branching fractions of $\rho^+ \to \pi^+ \pi^0$ or $K^{*+} \to K^+ \pi^0$ with $\pi^0 \to \gamma \gamma$.
The ST yield for the tag mode $i$ is $N_{{\rm ST}}^i$, and
the efficiency $\epsilon_{{\rm sig}}$ for detecting the signal $D^+$ decay
is averaged over the tag modes $i$,
\begin{equation}
  \label{br}
  {
    {\mathcal \epsilon}_{\rm sig} = \frac{\sum_i (N^i_{\rm ST}\cdot \epsilon^i_{\rm DT}/\epsilon^i_{\rm ST})}{N^{\rm tot}_{\rm ST}},}
\end{equation}
where $\epsilon^i_{\rm ST}$ is the efficiency of reconstructing the ST mode $i$ (referred to as the ST efficiency),
and $\epsilon^i_{ \rm DT}$ is the efficiency of finding the ST mode $i$ and the $D^{+} \to \gamma \rho^+$ or $D^{+} \to \gamma K^{*+}$ decay simultaneously (referred to as the DT efficiency).

\subsection{Single Tag Selection}

Charged tracks detected in the MDC (except for those used for $K^0_S$ reconstruction) are required to originate from a region within $|\rm{cos\theta}|<0.93$, $|V_{xy}|<$ 1\,cm, and $|V_{z}|<$ 10\,cm.
Here, $\theta$ is the polar angle of the charged track with respect to the MDC axis, $|V_{xy}|$ and $|V_{z}|$ are the distances of closest approach of the charged track to the interaction point perpendicular to and along the MDC axis, respectively.
Particle identification~(PID) for charged tracks combines measurements of the energy deposited in the MDC~(d$E$/d$x$) and the flight time in the TOF to form likelihoods $\mathcal{L}(h)~(h=\pi,K)$ for each hadron $h$ hypothesis.
Tracks are identified as charged kaons and pions by comparing the likelihoods for the kaon and pion hypotheses, requiring $\mathcal{L}(K)>\mathcal{L}(\pi)$ and $\mathcal{L}(\pi)>\mathcal{L}(K)$.

Each $K_{S}^0$ candidate is reconstructed from two oppositely charged tracks satisfying $|V_{z}|<$ 20~cm.
The two charged tracks are assigned as $\pi^+\pi^-$ without imposing further PID criteria.
They are constrained to originate from a common vertex and are required to have an invariant mass within (0.487, 0.511)~GeV$/c^{2}$.
The decay length of the $K^0_S$ candidate is required to be greater than twice the vertex resolution away from the interaction point.
The $\chi^2$ of the vertex fits (primary and secondary vertex fit) is required to be less than 100.

Photon candidates are selected by using the information recorded by the EMC. The time information of the crystal with the largest energy deposit inside a cluster is required to be within 700\,ns of the event start time.
The shower energy is required to be greater than $25~\mathrm{MeV}$ in the barrel $(|\cos \theta|<0.8)$ and $50~\mathrm{MeV}$ in the end cap $(0.86<|\cos \theta|<0.92)$ region.
The opening angle between the shower direction and the extrapolated position on the EMC of the closest  charged track must be greater than $10^{\circ}$.
The $\pi^0$ candidates are formed from photon pairs with an invariant mass within $(0.115,\,0.150)$\,GeV$/c^{2}$.
To improve the resolution, a kinematic fit constraining the $\gamma\gamma$ invariant mass to the known $\pi^{0}$ mass~\cite{ParticleDataGroup:2024cfk} is imposed on the selected photon pair.
The $\chi^2$ of the kinematic fit is required to be less than 50.
The four-momentum of the $\pi^0$ candidate updated by this kinematic fit is used for further analysis.

To separate $D^-$ mesons from combinatorial backgrounds, the energy difference $\Delta E_{\rm tag}$ is defined as $\Delta E_{\rm tag}\equiv E_{D^-}-E_{\mathrm{beam}}$, and the beam-constrained mass $M^{\rm tag}_{\rm BC}$ is defined as $M^{\rm tag}_{\rm BC}\equiv\sqrt{E_{\mathrm{beam}}^{2}/c^{4}-|\vec{p}_{D^-}|^{2}/c^{2}}$, where $E_{\mathrm{beam}}$ is the beam energy, and $E_{D^-}$ and $\vec{p}_{D^-}$ are the total energy and momentum of the $D^-$ candidate in the $e^+e^-$ center-of-mass frame, respectively.
If there is more than one $D^-$ candidate in a given ST mode, the candidate with the smallest value of $|\Delta E_{\rm tag}|$ will be kept for the subsequent analysis.
The $\Delta E_{\rm tag}$ requirements and ST efficiencies are listed in Table~\ref{ST:realdata}.

\begin{figure}[htbp]\centering
  \includegraphics[width=1.0\linewidth]{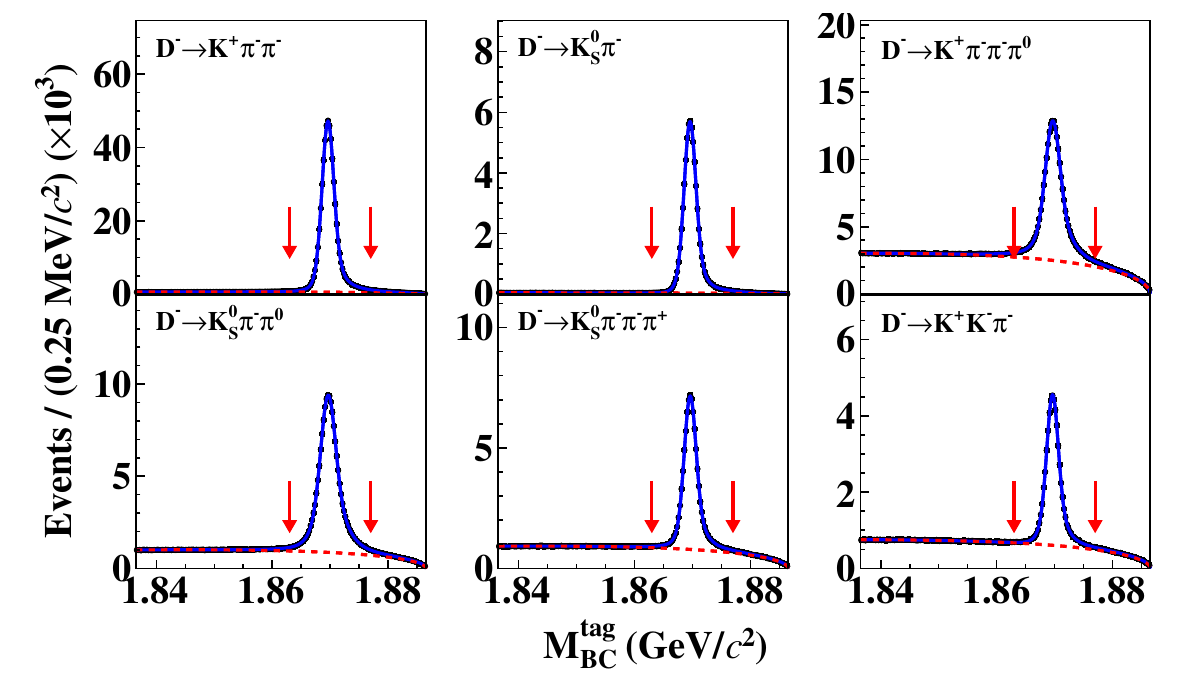}
  \caption{
    Fits to the $M^{\rm tag}_{\rm BC}$ distributions of the ST $D^-$ candidates for different tag modes.
    In each plot, the points with error bars are data, the blue curves are the best fits, and the red dashed curves describe the fitted combinatorial background shapes.
    The pair of red arrows indicates the signal window.}
    \label{fig:datafit_Massbc}
\end{figure}

The ST yield for each ST mode is extracted by performing an unbinned maximum likelihood fit to the corresponding $M^{\rm tag}_{\rm BC}$ distribution.
In the fit, the signal shape is derived from the MC-simulated signal shape convolved with a double-Gaussian function to compensate for the resolution difference between data and MC simulation.
The central values and resolutions of the double-Gaussian are both on the order of one per mille.
The combinatorial background shape is described by the ARGUS function~\cite{ARGUS:1990hfq}, with the end-point parameter fixed at $E_{\rm beam}=1.8865$~GeV/$c^{2}$.
Small peaking backgrounds for the ST modes $D^-\to K^0_S\pi^-$ and $D^-\to K^0_S\pi^+\pi^-\pi^-$ (with fractions of $<0.2\%$), estimated with the inclusive MC sample, have been subtracted away from the corresponding ST yields in data and ST efficiencies; while the peaking backgrounds for other four ST modes are negligible.
Figure~\ref{fig:datafit_Massbc} shows the fits to the $M^{\rm tag}_{\rm BC}$ distributions of the accepted ST candidates in data for different ST modes. The candidates with $M^{\rm tag}_{\rm BC}$ within $(1.863,1.877)$ GeV/$c^2$ are kept for further analyses. Summing over the tag modes, we obtain the total yield of ST $D^-$ mesons to be $(10638.3\pm3.6_{\rm stat})\times 10^3$ events.

\begin{table*}
  \renewcommand{\arraystretch}{1.2}
  \centering
  \caption {The $\Delta E_{\rm tag}$ requirements, the ST $D^-$ yields in data~($N^i_{\rm ST}$), the ST efficiencies ($\epsilon_{\rm ST}^{i}$), and the DT efficiencies ($\epsilon_{\rm DT}^{i}$) tagged $D^+ \to \gamma \rho^+$ and $D^+ \to \gamma K^{*+}$ for each tag modes. The uncertainties are statistical only.}
  \renewcommand\arraystretch{1.2}
    \begin{tabular}{cccccc}
      \hline
      \hline
      Tag mode                                & $\Delta E_{\rm tag}$~(MeV) & $N^{i}_{\rm ST}~(\times 10^3)$ & $\epsilon^i_{\rm ST}~(\%)$ & $\epsilon^{ia}_{\rm DT}~(\%)$ & $\epsilon^{ib}_{\rm DT}~(\%)$ \\\hline
$D^- \to  K^{+} \pi^{-} \pi^{-}$                  	 &$(-25,24)$	 &5567.2$\pm$2.5	 &51.08$\pm$0.01	 &9.90$\pm$0.09	 &8.96$\pm$0.09\\
$D^- \to  K^{+} \pi^{-} \pi^{-}  \pi^{0}$         	 &$(-57,46)$	 &1740.2$\pm$1.9	 &24.53$\pm$0.01	 &4.27$\pm$0.06	 &3.81$\pm$0.06\\
$D^- \to  K^{+}  K^{-} \pi^{-}$                   	 &$(-24,23)$	 &481.4$\pm$0.9	 &40.91$\pm$0.01	 &7.07$\pm$0.08	 &6.47$\pm$0.08\\
$D^- \to  K_{S}^{0} \pi^{-}$                      	 &$(-25,26)$	 &656.5$\pm$0.8	 &51.42$\pm$0.01	 &10.59$\pm$0.10	 &9.55$\pm$0.09\\
$D^- \to  K_{S}^{0} \pi^{-} \pi^{0}$              	 &$(-62,49)$	 &1442.4$\pm$1.5	 &26.45$\pm$0.01	 &5.10$\pm$0.07	 &4.50$\pm$0.07\\
$D^- \to  K_{S}^{0} \pi^{-} \pi^{-}  \pi^{+}$     	 &$(-28,27)$	 &790.2$\pm$1.1	 &29.68$\pm$0.01	 &5.32$\pm$0.07	 &4.71$\pm$0.07\\
\hline
\hline
\end{tabular}
\label{ST:realdata}
\end{table*}

\subsection{Double Tag Selection}

The candidates for $D^+ \to \gamma \rho^+$ and $D^+ \to \gamma K^{*+}$ are selected from the remaining tracks in presence of the tagged $D^-$ candidates.
The $\rho^+$ and $K^{*+}$ are reconstructed via $\rho^+ \to \pi^+ \pi^0$ and $K^{*+} \to K^+ \pi^0$, respectively.
Candidates for $\pi^+$, $K^+$, $\pi^0$, and $\gamma$ are selected with the same criteria as those used in the tag selection.
To select the candidates for $D^+\to\gamma \rho^+$ and $D^+\to\gamma K^{*+}$, the photon with the highest energy in a given event is selected as the radiative photon.
To suppress the backgrounds containing extra $\pi^0$ mesons, we require that there are no additional combinations of two photons~($N^{\pi^0}_{\rm extra}=0$) that satisfy the $\pi^0$ selection criteria in the event selection.
We require that there is no extra charged track~($N^{\rm charge}_{\rm extra}=0$) reconstructed in the signal decay.
The invariant mass requirements of the vector meson~($M_{V^+}$, where $V$ donates $\rho^+$ or $K^{*+}$), are optimized based on the Punzi figure-of-merit $\epsilon/(1.5+\sqrt{B})$~\cite{Punzi:2020fsv}, where $\epsilon$ is the signal efficiency based on the exclusive MC sample and $B$ is the background yield obtained from the inclusive MC sample.
According to the optimization, we require the candidates for $D^+\to\gamma \rho^+$ and $D^+\to\gamma K^{*+}$ to
satisfy $M_{\pi^+ \pi^0}\in (0.64,0.89)$ GeV/$c^2$ and $M_{K^+ \pi^0}\in (0.84,0.94)$ $\mathrm{GeV}/c^2$, respectively.
We define the missing mass squared of the
radiative photon
  $M^{2}_{\gamma} \equiv (\sqrt{s} -\Sigma_k E_k)^2 - \left| \Sigma_k \vec{p}_{k} \right|^{2}$,
in which $E_k$ and $\vec p_k$ are the energy and momentum of the ST $D^-$ or $V^+$, respectively.
To remove the background from $D^{+} \to K^0_L V^+$, which peaks in the $M^{2}_{\gamma}$ distribution around 0.25~GeV$^2$/$c^4$ corresponding to the square of the known $K^0_L$ mass~\cite{ParticleDataGroup:2024cfk}, the value of $M^{2}_{\gamma}$ is required to be less than 0.07~GeV$^2$/$c^4$ based on the Punzi optimization.
To extract the information of the signal side, we define two kinematic variables of
$M_{\rm BC}^{\rm sig}$ and $\Delta E_{\rm sig}$ similarly as in the tag side.
After the optimization, we require the candidate events to satisfy $M_{\rm BC}^{\rm sig}\in (1.865,1.873)~\mathrm{GeV}/c^2$.

After applying the above requirements, the average signal efficiencies in the presence of the ST $D^-$ mesons are $(18.92\pm0.11)\%$ and $(17.02\pm0.11)\%$ for $D^+\to\gamma \rho^+$ and $D^+\to\gamma K^{*+}$, respectively. These efficiencies do not include the branching fractions of subdecays.

\section{Results}

\begin{figure}[htpb]
  \centering
  \includegraphics[width=1.0\linewidth]{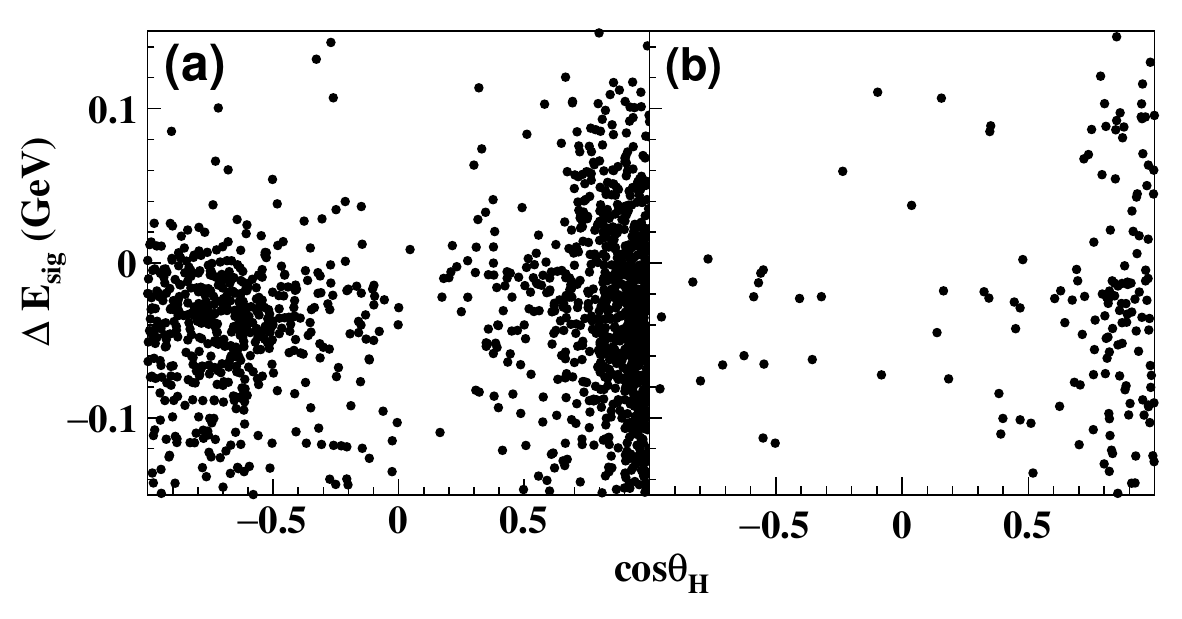}
  \caption{
    Distributions of $\Delta E_{\rm sig}$ versus $\cos \theta_{\rm H}$ of the DT candidate events for $D^+\to\gamma \rho^+$(a) and $D^+\to\gamma K^{*+}$(b) in data.
    \label{plot:helix2d}
  }
\end{figure}

\begin{figure}[htpb]
  \centering
  \includegraphics[width=1.0\linewidth]{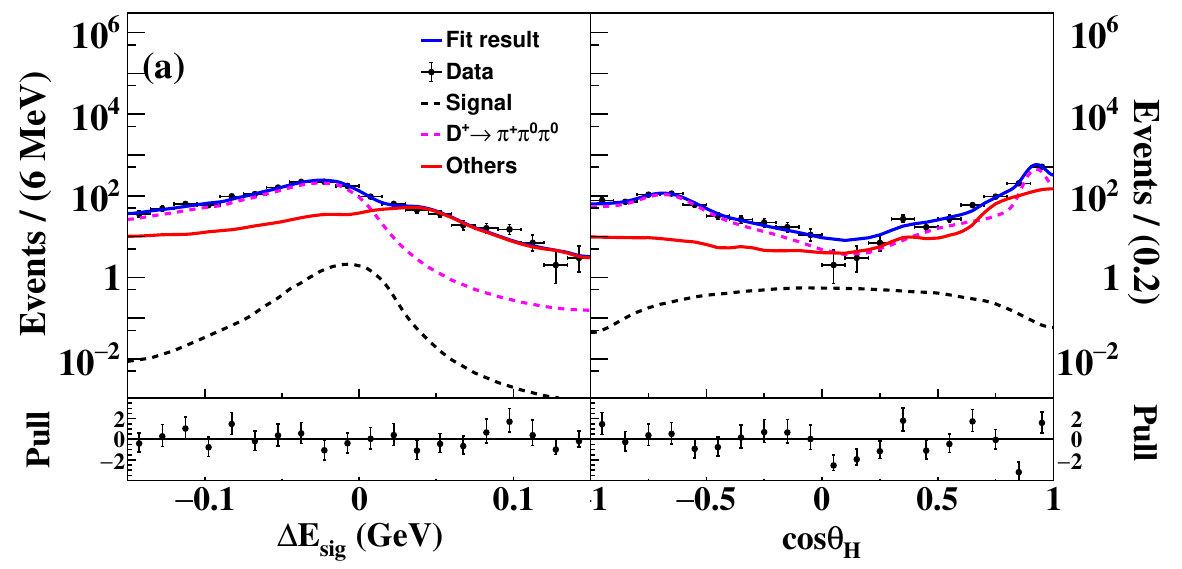}
  \includegraphics[width=1.0\linewidth]{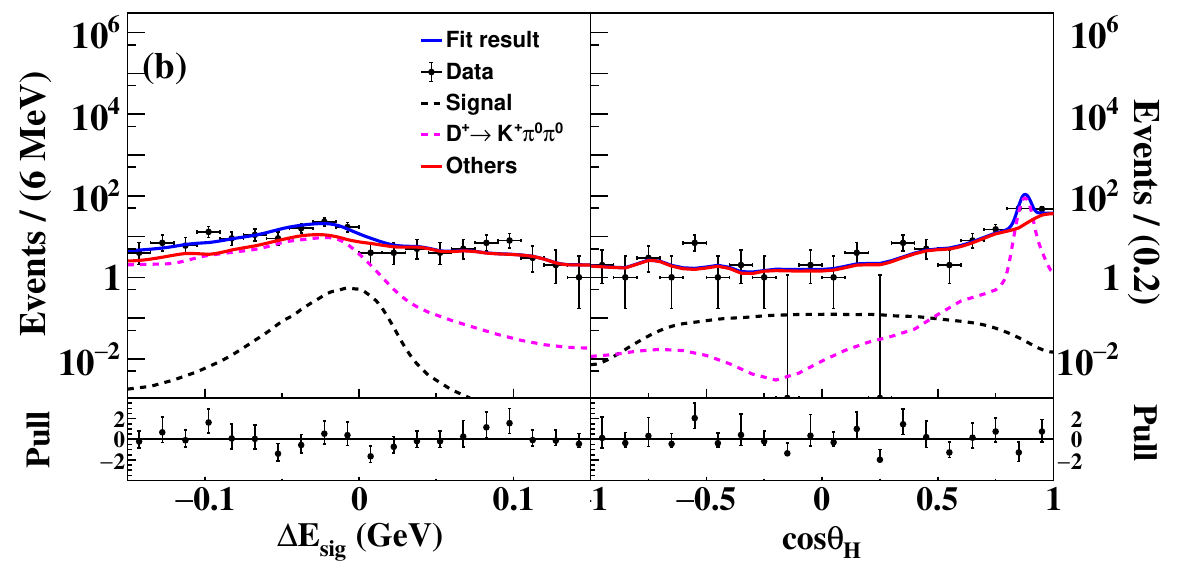}
  \caption{
    Projections on $\Delta E_{\rm sig}$ versus $\cos \theta_{\rm H}$ of the 2D fit for the DT candidate events of $D^+\to\gamma \rho^+$ (a) and $D^+ \to \gamma K^{*+}$ (b).
    The dots with error bars correspond to
    data.
    The blue solid curves correspond to
    the fit results.
    The black dashed lines correspond to
    the fitted signal, the pink dashed curves correspond to
    the dominant background contributions, and the red solid curves correspond to
    other background contributions.}
  \label{2Dfit1}
\end{figure}

To further study combinatorial backgrounds and extract the signal yields, we define the helicity angle $\theta_{\rm H}$ of $\pi^+(K^+)$, which is the angle between the $\pi^+(K^+)$ momentum in the $\rho^+(K^{*+})$ rest frame and $\rho^+(K^{*+})$ momentum in the $D^+$ rest frame.
The $\Delta E_{\rm sig}$ and $M_{\rm BC}^{\rm sig}$ ensure the property of a $D$ meson, $\cos\theta_{H}$ reveals that of the $D\to\gamma V$ decay.
The two-dimensional distributions of the $\Delta E_{\rm sig}$ versus $\cos \theta_{\rm H}$ of $\pi^+(K^+)$ in data are shown in Fig.~\ref{plot:helix2d}.
The signal shape in the $\cos\theta_{H}$ distribution is arched, while the corresponding background shape is concave, as shown in Fig.\ref{2Dfit1}.

\begin{figure*}[htbp]
  \centering
  \includegraphics[width=0.95\textwidth]{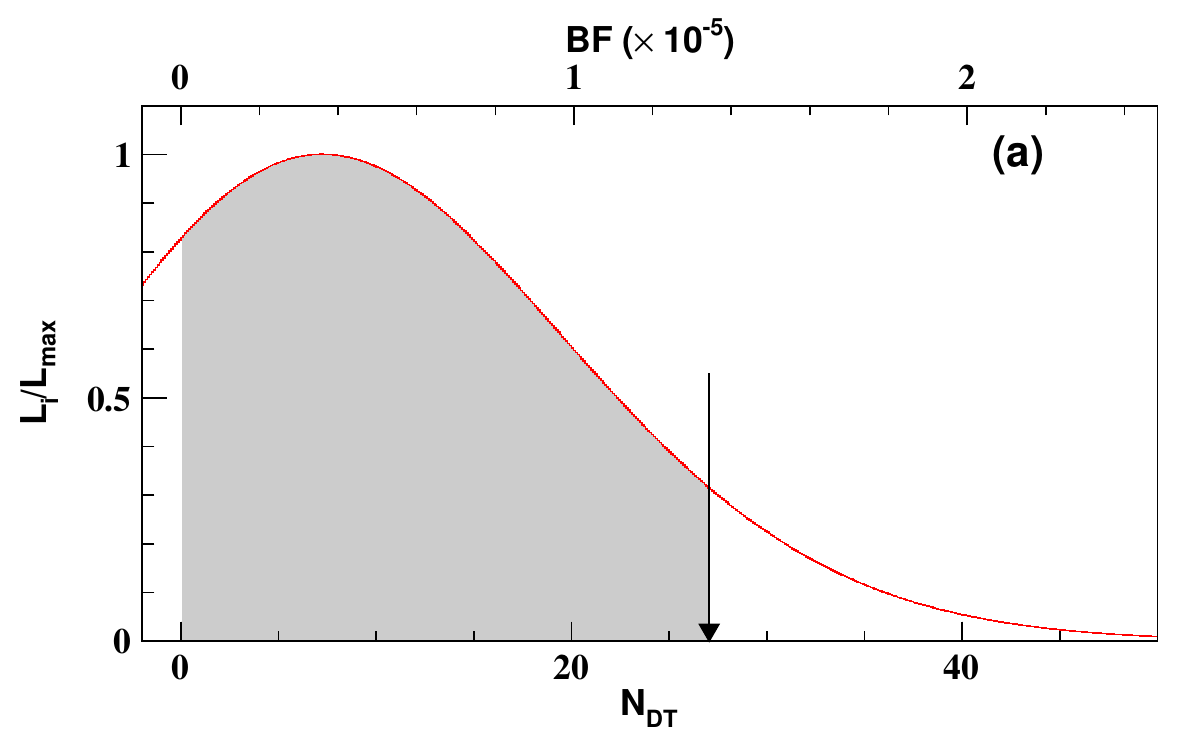}
  \includegraphics[width=0.95\textwidth]{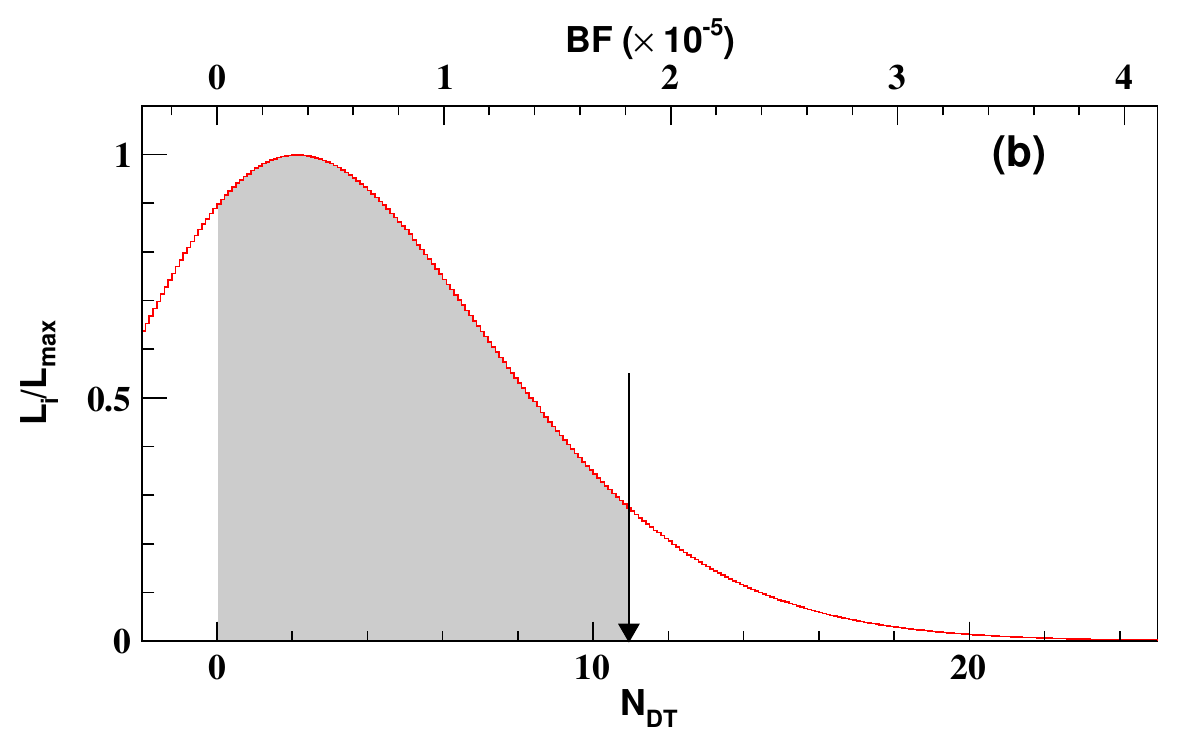}
  \caption{Distributions of normalized likelihood versus the signal yield $N_{\rm DT}$ or normalized branching fraction of $D^+\to\gamma \rho^+$(a) and $D^+\to\gamma K^{*+}$(b).
  The results obtained with the systematic uncertainties incorporated  are shown by the red curves. The black arrow shows the result corresponding to 90\% confidence level.}
  \label{fig:upper}
\end{figure*}

The signal yields are extracted from a 2D unbinned maximum likelihood fit to the $\Delta E_{\rm sig}$ versus $\cos \theta_{\rm H}$ distribution for $D^+\to\gamma \rho^+$ and $D^+\to\gamma K^{*+}$, respectively.
In the fits, the signal shapes are derived from the signal MC sample,
and the background shapes are derived from the inclusive MC sample with the RooKeysPDF tool~\cite{ROOTonline}.
The signal and background yields are both allowed to float in the fit.
The yields of the dominated background, $D^+\to\pi^+\pi^0\pi^0$ or $D^+\to K^+\pi^0\pi^0$, are fixed by the mis-identification and the branching fractions.
Figure~\ref{2Dfit1} shows the fitted results for $D^+\to\gamma \rho^+$ and $D^+\to\gamma K^{*+}$.
The signal yields are
$6.8^{+12.4}_{-11.2}$ for $D^+\to\gamma \rho^+$
and
$1.8^{+5.2}_{-4.3}$ for $D^+\to\gamma K^{*+}$.
Since no significant signal is found,
we set the upper limits as shown in Fig.~\ref{fig:upper} by using the Bayesian approach~\cite{Feldman:1997qc,Stenson:2006gwf,Convery:2003af,BESIII:2021drk} after considering the systematic uncertainties discussed later.
Finally, the upper limits on the branching fractions of $D^+ \to \gamma \rho^+$ and $D^+ \to \gamma K^{*+}$ at 90\% confidence level are set to be $1.3\times10^{-5}$ and $1.8\times10^{-5}$, respectively.

\section{Systematic uncertainty}

With the DT method, many systematic uncertainties associated with the ST selection cancel and do not affect the branching fraction measurement.
This concerns tracking, PID, subdecay BFs, and the $\Delta E$ requirements in the ST selection.

To account for the additive systematic uncertainties related to the fits,
the alternative fits involve different RooKeysPDF smooth parameters by varying $\pm 0.5$ from the default value of 1.0.
Among the results of these fits, the largest upper limit on the branching fraction is chosen.
The multiplicative systematic uncertainties are discussed below.

The uncertainty associated with the ST yield $N_{\rm ST}^{\rm tot}$ is assigned as 0.1\% after varying the signal and background shapes and floating the parameters of one Gaussian in the fit.
The tracking and PID efficiencies of $K^\pm$ and $\pi^\pm$ are studied by analyzing DT $D^{0}\bar{D}^0(D^+D^-)$ events, where the control samples comprise hadronic decays of $D^{0}\to K^-\pi^+$, $D^{0}\to K^- \pi^+\pi^0$, $D^{0}\to K^-\pi^+\pi^+\pi^-$ versus $D^0 \to K^+\pi^-$, $D^{0}\to K^+ \pi^-\pi^0$, $D^{0}\to K^+\pi^-\pi^-\pi^+$ as well as $D^+ \to K^-\pi^+\pi^+$ versus $D^+ \to K^+\pi^-\pi^-$.
The systematic uncertainty due to tracking efficiencies is assigned as 1.0\% for both $K^\pm$ and $\pi^\pm$;
the systematic uncertainty due to PID is assigned as 1.0\% for $K^\pm$ and $\pi^\pm$.
The systematic uncertainty related to the $\gamma$ selection is studied with the $J/\psi\to \pi^+\pi^-\pi^0$ decay~\cite{BESIII:2011ysp}, while the systematic uncertainty on the $\pi^0$ selection is examined through the DT hadronic $D\bar D$ events, as in Ref.~\cite{BESIII:2018mwk}.
These uncertainties are assigned as 1.0\% per photon and 2.0\% per $\pi^0$.

The systematic uncertainties due to the $\rho^+$ and $K^{*+}$ mass windows are estimated by using the DT samples of
$\bar D^0\to K^+ \pi^- $, $\bar D^0\to K^+ \pi^- \pi^0$, $\bar D^0\to K^+ \pi^- \pi^- \pi^+$ versus $D^0\to \pi^- \pi^0 e^+ \nu_e$ and $D^0\to K^- \pi^0 e^+ \nu_e$, respectively.
The selection criteria of candidates are the same as in Ref.~\cite{BESIII:2023exq}.
The differences of the acceptance efficiencies between data and MC simulation, 0.5\% and  0.1\%, are assigned as the systematic uncertainties for $D^+\to\gamma \rho^+$ and $D^+\to\gamma K^{*+}$, respectively.

The differences in the $M^{\rm sig}_{\rm BC}$ resolution between data and MC simulation are obtained from the DT events of $D^+ \to \pi^+\pi^0\pi^0$ reconstructed versus the same tag modes used in the baseline analysis. The discrepancy in acceptance efficiencies between data and MC simulation of 0.1\% is taken as the related systematic uncertainty.

The combined systematic uncertainty from the $N^{\pi^0}_{\rm extra}$ and $ N^{\rm charge}_{\rm extra}$ requirements is estimated to be 0.7\%, which is also assigned by analyzing the control sample of $D^+\to \pi^+\pi^0\pi^0$.

The systematic uncertainty from the $M^{2}_{\gamma}$ requirement is estimated to be 0.6\%, with the control sample of $D^+\to \pi^+\pi^0\pi^0$ by missing a $\gamma$ in the final state.

The uncertainties due to the limited size of MC samples are 0.6\% and 0.7\% for $D^+\to\gamma \rho^+$ and $D^+\to\gamma K^{*+}$, respectively.

The uncertainties of the branching fractions of $\rho^+ \to \pi^+ \pi^0$, $K^{*+} \to K^+ \pi^0$ and $\pi^0 \to \gamma \gamma$ are negligible~\cite{ParticleDataGroup:2024cfk}.

The total multiplicative systematic uncertainty is obtained by adding the individual components in quadrature.
Table~\ref{tab:relsysuncertainties} summarizes the sources of the systematic uncertainties in the branching fraction measurements.

\begin{table}[htpb]
  \centering
  \caption{
    Relative multiplicative systematic uncertainties (\%) in the branching fraction measurements.}
  \label{tab:relsysuncertainties}
  \centering
  \begin{tabular}{c|c|c}
    \hline \hline
    Source                                                       & $D^+ \to \gamma \rho^+$ & $D^+ \to \gamma K^{*+}$ \\
    \hline
    $N_{\rm ST}$                                                     & 0.1                     & 0.1                     \\
    Tracking                                                          & 1.0                     & 1.0                     \\
    PID                                                               & 1.0                     & 1.0                     \\
    $\gamma$ and $\pi^0$ selection                                    & 3.0                     & 3.0                     \\
    $M_{V^{+}}$ requirement                                           & 0.5                     & 0.1                     \\
    $M^{\rm sig}_{\rm BC}$ requirement                                          & 0.1                     & 0.1                     \\
    $N_{\rm extra}^{\rm charge} \& N^{\pi^0}_{\rm extra}$ requirement & 0.7                     & 0.7                     \\
    $M^{2}_{\gamma}$ requirement                          & 0.6                    & 0.6                     \\
    MC statistics                                                     & 0.6                     & 0.7                     \\
    \hline
    Total                                                             & 3.6                     & 3.5                     \\
    \hline \hline
  \end{tabular}
\end{table}

\section{Summary}

By analyzing a data sample corresponding to an integrated luminosity of 20.3~fb$^{-1}$ collected in $e^+e^-$ annihilations recorded at $\sqrt{s} = 3.773\;\text{GeV}$, we search for the radiative decays $D^{+} \to \gamma \rho^+$ and $D^{+} \to \gamma K^{*+}$. No significant signals are observed.
The upper limits on the branching fractions of $D^{+} \to \gamma \rho^+$ and $D^{+} \to \gamma K^{*+}$ at 90\% confidence level are set to be
$1.3\times10^{-5}$ and $1.8\times10^{-5}$, respectively.
Table~\ref{tab:theo} summarizes the different theoretically predicted BFs and our measured upper limits.
The results obtained in this analysis are compatible with theoretical calculations, except that the upper limit on the branching fraction of $D^+\to \gamma \rho^+$ deviates with theoretical calculation in the SM~\cite{Fu:2018yin} by about $3.6\sigma$.
The larger data sample at the Super Tau-Charm Facility~\cite{Achasov:2023gey} will offer an opportunity to further enhance the sensitivity of the search for these radiative decays.

\begin{table}[htpb]
  \centering
  \caption{The predicted branching fractions$~(\times 10^{-5})$ of $D^+\to \gamma \rho^+$ and $D^+ \to \gamma K^{*+}$ from various theories, and the upper limits in this work.}
    \begin{tabular}{c|c|c}
      \hline \hline
      Result                                     & $D^+ \to \gamma \rho^+$   & $D^+ \to \gamma K^{*+}$ \\ \hline
      PoleDiagram and VMD ~\cite{Burdman:1995te} & $2-6$                     & $0.1-0.3$               \\
      SM         ~\cite{Fu:2018yin}              & $5.0\pm0.9$ & --                      \\
      QCD SM       ~\cite{Khodjamirian:1995uc}   & $0.46$                    & --                      \\
      Hybrid       ~\cite{deBoer:2017que}        & $0.017 - 2.33$         & $0.048-0.76$                      \\
      FS         ~\cite{Fajfer:1997bh}           & $1.8 - 4.1$               & $0.25-0.5$         \\
      Factorization    ~\cite{Fajfer:1998dv}     & $0.4 - 6.3$               & $0.03 - 0.44$           \\
      \hline
      This work  & $<1.3$   &$<1.8$  \\
      \hline \hline
    \end{tabular}
  \label{tab:theo}
\end{table}

\section{Acknowledgement}
The BESIII Collaboration thanks the staff of BEPCII and the IHEP computing center for their strong support. This work is supported in part by National Key R\&D Program of China under Contracts Nos. 2023YFA1606000, 2020YFA0406300, 2020YFA0406400; National Natural Science Foundation of China (NSFC) under Contracts Nos. 11875170, 12105076, 11635010, 11735014, 11935015, 11935016, 11935018, 12025502, 12035009, 12035013, 12061131003, 12192260, 12192261, 12192262, 12192263, 12192264, 12192265, 12221005, 12225509, 12235017, 12361141819; the Chinese Academy of Sciences (CAS) Large-Scale Scientific Facility Program; the CAS Center for Excellence in Particle Physics (CCEPP); Joint Large-Scale Scientific Facility Funds of the NSFC and CAS under Contract No. U1832207; 100 Talents Program of CAS; The Institute of Nuclear and Particle Physics (INPAC) and Shanghai Key Laboratory for Particle Physics and Cosmology; German Research Foundation DFG under Contracts Nos. 455635585, FOR5327, GRK 2149; Istituto Nazionale di Fisica Nucleare, Italy; Ministry of Development of Turkey under Contract No. DPT2006K-120470; National Research Foundation of Korea under Contract No. NRF-2022R1A2C1092335; National Science and Technology fund of Mongolia; National Science Research and Innovation Fund (NSRF) via the Program Management Unit for Human Resources \& Institutional Development, Research and Innovation of Thailand under Contract No. B16F640076; Polish National Science Centre under Contract No. 2019/35/O/ST2/02907; The Swedish Research Council; U. S. Department of Energy under Contract No. DE-FG02-05ER41374

\bibliographystyle{JHEP}
\bibliography{refer}

\clearpage

M.~Ablikim$^{1}$, M.~N.~Achasov$^{4,c}$, P.~Adlarson$^{76}$, O.~Afedulidis$^{3}$, X.~C.~Ai$^{81}$, R.~Aliberti$^{35}$, A.~Amoroso$^{75A,75C}$, Q.~An$^{72,58,a}$, Y.~Bai$^{57}$, O.~Bakina$^{36}$
% ...
\\

{\it
~\\
$^{1}$ Institute of High Energy Physics, Beijing 100049, People's Republic of China\\
$^{2}$ Beihang University, Beijing 100191, People's Republic of China\\
$^{3}$ Bochum  Ruhr-University, D-44780 Bochum, Germany\\
$^{4}$ Budker Institute of Nuclear Physics SB RAS (BINP), Novosibirsk 630090, Russia\\
$^{5}$ Carnegie Mellon University, Pittsburgh, Pennsylvania 15213, USA\\
$^{6}$ Central China Normal University, Wuhan 430079, People's Republic of China\\
$^{7}$ Central South University, Changsha 410083, People's Republic of China\\
$^{8}$ China Center of Advanced Science and Technology, Beijing 100190, People's Republic of China\\
$^{9}$ China University of Geosciences, Wuhan 430074, People's Republic of China\\
$^{10}$ Chung-Ang University, Seoul, 06974, Republic of Korea\\
$^{11}$ COMSATS University Islamabad, Lahore Campus, Defence Road, Off Raiwind Road, 54000 Lahore, Pakistan\\
$^{12}$ Fudan University, Shanghai 200433, People's Republic of China\\
$^{13}$ GSI Helmholtzcentre for Heavy Ion Research GmbH, D-64291 Darmstadt, Germany\\
$^{14}$ Guangxi Normal University, Guilin 541004, People's Republic of China\\
$^{15}$ Guangxi University, Nanning 530004, People's Republic of China\\
$^{16}$ Hangzhou Normal University, Hangzhou 310036, People's Republic of China\\
$^{17}$ Hebei University, Baoding 071002, People's Republic of China\\
$^{18}$ Helmholtz Institute Mainz, Staudinger Weg 18, D-55099 Mainz, Germany\\
$^{19}$ Henan Normal University, Xinxiang 453007, People's Republic of China\\
$^{20}$ Henan University, Kaifeng 475004, People's Republic of China\\
$^{21}$ Henan University of Science and Technology, Luoyang 471003, People's Republic of China\\
$^{22}$ Henan University of Technology, Zhengzhou 450001, People's Republic of China\\
$^{23}$ Huangshan College, Huangshan  245000, People's Republic of China\\
$^{24}$ Hunan Normal University, Changsha 410081, People's Republic of China\\
$^{25}$ Hunan University, Changsha 410082, People's Republic of China\\
$^{26}$ Indian Institute of Technology Madras, Chennai 600036, India\\
$^{27}$ Indiana University, Bloomington, Indiana 47405, USA\\
$^{28}$ INFN Laboratori Nazionali di Frascati , (A)INFN Laboratori Nazionali di Frascati, I-00044, Frascati, Italy; (B)INFN Sezione di  Perugia, I-06100, Perugia, Italy; (C)University of Perugia, I-06100, Perugia, Italy\\
$^{29}$ INFN Sezione di Ferrara, (A)INFN Sezione di Ferrara, I-44122, Ferrara, Italy; (B)University of Ferrara,  I-44122, Ferrara, Italy\\
$^{30}$ Inner Mongolia University, Hohhot 010021, People's Republic of China\\
$^{31}$ Institute of Modern Physics, Lanzhou 730000, People's Republic of China\\
$^{32}$ Institute of Physics and Technology, Peace Avenue 54B, Ulaanbaatar 13330, Mongolia\\
$^{33}$ Instituto de Alta Investigaci\'on, Universidad de Tarapac\'a, Casilla 7D, Arica 1000000, Chile\\
$^{34}$ Jilin University, Changchun 130012, People's Republic of China\\
$^{35}$ Johannes Gutenberg University of Mainz, Johann-Joachim-Becher-Weg 45, D-55099 Mainz, Germany\\
$^{36}$ Joint Institute for Nuclear Research, 141980 Dubna, Moscow region, Russia\\
$^{37}$ Justus-Liebig-Universitaet Giessen, II. Physikalisches Institut, Heinrich-Buff-Ring 16, D-35392 Giessen, Germany\\
$^{38}$ Lanzhou University, Lanzhou 730000, People's Republic of China\\
$^{39}$ Liaoning Normal University, Dalian 116029, People's Republic of China\\
$^{40}$ Liaoning University, Shenyang 110036, People's Republic of China\\
$^{41}$ Nanjing Normal University, Nanjing 210023, People's Republic of China\\
$^{42}$ Nanjing University, Nanjing 210093, People's Republic of China\\
$^{43}$ Nankai University, Tianjin 300071, People's Republic of China\\
$^{44}$ National Centre for Nuclear Research, Warsaw 02-093, Poland\\
$^{45}$ North China Electric Power University, Beijing 102206, People's Republic of China\\
$^{46}$ Peking University, Beijing 100871, People's Republic of China\\
$^{47}$ Qufu Normal University, Qufu 273165, People's Republic of China\\
$^{48}$ Renmin University of China, Beijing 100872, People's Republic of China\\
$^{49}$ Shandong Normal University, Jinan 250014, People's Republic of China\\
$^{50}$ Shandong University, Jinan 250100, People's Republic of China\\
$^{51}$ Shanghai Jiao Tong University, Shanghai 200240,  People's Republic of China\\
$^{52}$ Shanxi Normal University, Linfen 041004, People's Republic of China\\
$^{53}$ Shanxi University, Taiyuan 030006, People's Republic of China\\
$^{54}$ Sichuan University, Chengdu 610064, People's Republic of China\\
$^{55}$ Soochow University, Suzhou 215006, People's Republic of China\\
$^{56}$ South China Normal University, Guangzhou 510006, People's Republic of China\\
$^{57}$ Southeast University, Nanjing 211100, People's Republic of China\\
$^{58}$ State Key Laboratory of Particle Detection and Electronics, Beijing 100049, Hefei 230026, People's Republic of China\\
$^{59}$ Sun Yat-Sen University, Guangzhou 510275, People's Republic of China\\
$^{60}$ Suranaree University of Technology, University Avenue 111, Nakhon Ratchasima 30000, Thailand\\
$^{61}$ Tsinghua University, Beijing 100084, People's Republic of China\\
$^{62}$ Turkish Accelerator Center Particle Factory Group, (A)Istinye University, 34010, Istanbul, Turkey; (B)Near East University, Nicosia, North Cyprus, 99138, Mersin 10, Turkey\\
$^{63}$ University of Bristol, (A)H H Wills Physics Laboratory; (B)Tyndall Avenue; (C)Bristol; (D)BS8 1TL\\
$^{64}$ University of Chinese Academy of Sciences, Beijing 100049, People's Republic of China\\
$^{65}$ University of Groningen, NL-9747 AA Groningen, The Netherlands\\
$^{66}$ University of Hawaii, Honolulu, Hawaii 96822, USA\\
$^{67}$ University of Jinan, Jinan 250022, People's Republic of China\\
$^{68}$ University of Manchester, Oxford Road, Manchester, M13 9PL, United Kingdom\\
$^{69}$ University of Muenster, Wilhelm-Klemm-Strasse 9, 48149 Muenster, Germany\\
$^{70}$ University of Oxford, Keble Road, Oxford OX13RH, United Kingdom\\
$^{71}$ University of Science and Technology Liaoning, Anshan 114051, People's Republic of China\\
$^{72}$ University of Science and Technology of China, Hefei 230026, People's Republic of China\\
$^{73}$ University of South China, Hengyang 421001, People's Republic of China\\
$^{74}$ University of the Punjab, Lahore-54590, Pakistan\\
$^{75}$ University of Turin and INFN, (A)University of Turin, I-10125, Turin, Italy; (B)University of Eastern Piedmont, I-15121, Alessandria, Italy; (C)INFN, I-10125, Turin, Italy\\
$^{76}$ Uppsala University, Box 516, SE-75120 Uppsala, Sweden\\
$^{77}$ Wuhan University, Wuhan 430072, People's Republic of China\\
$^{78}$ Yantai University, Yantai 264005, People's Republic of China\\
$^{79}$ Yunnan University, Kunming 650500, People's Republic of China\\
$^{80}$ Zhejiang University, Hangzhou 310027, People's Republic of China\\
$^{81}$ Zhengzhou University, Zhengzhou 450001, People's Republic of China\\
~\\
}
~\\
{
    $^{a}$ Deceased\\
    $^{b}$ Also at the Moscow Institute of Physics and Technology, Moscow 141700, Russia\\
    $^{c}$ Also at the Novosibirsk State University, Novosibirsk, 630090, Russia\\
    $^{d}$ Also at the NRC "Kurchatov Institute", PNPI, 188300, Gatchina, Russia\\
    $^{e}$ Also at Goethe University Frankfurt, 60323 Frankfurt am Main, Germany\\
    $^{f}$ Also at Key Laboratory for Particle Physics, Astrophysics and Cosmology, Ministry of Education; Shanghai Key Laboratory for Particle Physics and Cosmology; Institute of Nuclear and Particle Physics, Shanghai 200240, People's Republic of China\\
    $^{g}$ Also at Key Laboratory of Nuclear Physics and Ion-beam Application (MOE) and Institute of Modern Physics, Fudan University, Shanghai 200443, People's Republic of China\\
    $^{h}$ Also at State Key Laboratory of Nuclear Physics and Technology, Peking University, Beijing 100871, People's Republic of China\\
    $^{i}$ Also at School of Physics and Electronics, Hunan University, Changsha 410082, China\\
    $^{j}$ Also at Guangdong Provincial Key Laboratory of Nuclear Science, Institute of Quantum Matter, South China Normal University, Guangzhou 510006, China\\
    $^{k}$ Also at MOE Frontiers Science Center for Rare Isotopes, Lanzhou University, Lanzhou 730000, People's Republic of China\\
    $^{l}$ Also at Lanzhou Center for Theoretical Physics, Lanzhou University, Lanzhou 730000, People's Republic of China\\
    $^{m}$ Also at the Department of Mathematical Sciences, IBA, Karachi 75270, Pakistan\\
    $^{n}$ Also at Ecole Polytechnique Federale de Lausanne (EPFL), CH-1015 Lausanne, Switzerland\\
    $^{o}$ Also at Helmholtz Institute Mainz, Staudinger Weg 18, D-55099 Mainz, Germany\\
    $^{p}$ Also at School of Physics, Beihang University, Beijing 100191 , China\\
}

\end{document}